\documentclass[a4paper]{svmult}
\usepackage{times}
\usepackage[dvips]{graphicx,epsfig}

\usepackage{amsmath,amsfonts,amssymb}

\usepackage{cite,url}

\begin{document}
%%%

\title*{Resonance contribution to single pion production in neutrino-nucleon scattering}
\author{K.M. Graczyk}

\titlerunning{Resonance contribution to single pion production in neutrino-nucleon scattering}
\authorrunning{{K.M. Graczyk}}

\toctitle{Resonance contribution to single pion production in neutrino-nucleon scattering}
\tocauthor{{K.M. Graczyk}}

\institute{{Dipartimento di Fisica Teorica, University of Torino and INFN,
Sezione di Torino, via P. Giuria 1, I-10125 Torino, Italy} \and { Institute of Theoretical Physics, University of Wroc\l
aw, pl. M. Borna 9, 50-204, Wroc\l aw, Poland}
}

\maketitle

\begin{abstract}
Single pion production in neutrino-nucleon scattering is discussed. The neutrino energies characteristic for T2K project are considered. Two new parameterizations of $C_5^A$ axial form factor are proposed. Both of them are obtained in simultaneous fit to ANL and BNL data. One of them (which fits better to BNL data) leads to $d\sigma/dQ^2$ differential cross section which is significantly reduced at low $Q^2$.
\end{abstract}

\section{Introduction}

Single pion production in neutrino-nucleon/nucleus scattering has been a subject of intensive theoretical and experimental studies for many years.  One of the first comprehensive theoretical model for  pion production in electron- and neutrino- nucleon scattering was proposed by Adler \cite{Adler}.  Since Adler model had appeared many other approaches have been proposed. These descriptions were mainly devoted to   electron-nucleon scattering. Some of them were extended to describe neutrino-nucleon interactions. Let us mention two of the newest
approaches: Sato and Lee model \cite{Sato:2003rq} and the model proposed by Hernandez \textit{et al.}\cite{Hernandez:2007qq}. Both of them are applied to describe charged and neutral current neutrino-nucleon interactions.

From the theoretical point of view there are useful  analogies  between electron-nucleon  and neutrino-nucleon interactions. The vector part of charged electro-weak hadronic current is related with the electromagnetic hadronic current -- it is a result of conserved vector current (CVC) hypothesis. For electron-nucleon scattering high accuracy experimental data are available. This allows to verify and fine tune theoretical approaches. Well tested models in electron-nucleon interaction should describe well the vector contribution to $\nu$-nucleon scattering, but for neutrino-nucleon scattering more important is the proper modeling of an axial contribution. The description of the axial contribution to the cross section can be mostly verified  by comparing with neutrino-nucleon scattering data.

Neutrino-matter interactions have been intensively studied for many years but experimental data for $1\pi$ production are limited. Most of the data sets have been collected for charged current $\nu N$ scattering, and there are only a few total cross section points for neutral current neutrino-nucleon scattering \cite{Krenz:1977sw}\cite{Derrick:1980nr}. A study of the axial contribution, a fitting of the axial form factors, requires analysis of $d\sigma/dQ^2$ differential cross sections. Such data have been  collected by several experiments. Let us mention two of  the most important experiments, namely the bubble chamber experiments,  ANL \cite{Radecky:1981fn} (Argone National Laboratory)  and BNL (Brookhaven National Laboratory) \cite{Kitagaki:1986ct}.

In the ANL experiment the differential cross section  $d\sigma/dQ^2$ was measured up to $Q^2 = 1$~GeV$^2$ while in BNL experiment data were collected up to $Q^2 = 3$~GeV$^2$. In both projects deuterium  was used as a target and very similar measurement techniques were applied, but the collected  cross section data  seem to be in a disagreement. The BNL total cross sections are systematically higher (by 20\%) than the ANL total cross sections \cite{Wascko:2006tx}. Additionally fitting to ANL $d\sigma/dQ^2$ data leads to a functional form of the axial form factor which has different $Q^2$ dependence than the one extracted from the BNL data \cite{Graczyk:2007bc}. Therefore, usually (in the literature) the axial form factor fits to either the ANL  or BNL data.

During the last few years the experiments K2K (KEK to Super-Kamiokande long-baseline neutrino oscillation experiment) \cite{K2K} and  MiniBooNE \cite{MiniBooNE} (Booster Neutrino Experiment at Fermilab)  have collected new data also for single pion production. Since in K2K  a water cherenkov detector (Super-Kamiokande) was used neutrino-oxygen interactions were mainly detected, while at MiniBooNE  neutrinos mainly interact with carbon nuclei.  Modeling of neutrino-nucleus scattering requires to consider some description of the  nucleus, which must be implemented in the data analysis.

K2K and MiniBooNE experiments are different in the design and they used different  measurement techniques  but  in both projects it has been observed that the $d\sigma/d Q^2$ data points at low $Q^2$ (four-momentum transfer) are below theoretical predictions  (Monte Carlo (MC) simulation) \cite{K2K_small_Q2}\cite{Miniboone_small_Q2}. This discrepancy between theory and experiment can be explained by assuming that the contribution from the coherent pion\footnote{In the coherent interaction the quantum numbers of the nucleon are unchanged and pions are produced mostly in the forward directions. } production  is negligible \cite{K2K_small_Q2}. On the other hand it is possible that this effect  can be explained by more careful discussion of nuclear effects. There are also possible other sources of this disagreement like improper account of the lepton mass effects \cite{Graczyk:2007xk}. Eventually, the $Q^2$ dependence of the cross section  is mainly constrained by  the axial form factor.

The investigation of $1\pi$ production in neutrino-nucleon scattering is important to understand the character of neutrino-matter interactions as well as the structure of the nucleons and nuclei. The subject is also important from the practical point of view: a proper prediction of cross sections for $1\pi$ production plays a crucial role in data analysis of long baseline oscillation experiments.  Indeed in K2K the oscillation of neutrinos was investigated by observing distortion of the $\nu_\mu$ energy spectrum in the far detector \cite{K2K_distortion}. The energy spectrum was reconstructed by considering quasi-elastic (QE) charged current events but some processes with single pion production stand  background to these events.

In few years T2K \cite{T2K} (Tokai to Kamioka), a next generation long baseline neutrino oscillation experiment,  will start collecting data. A study of $\nu_\mu\to\nu_{e}$ oscillation is the main goal of the project.  It can be only investigated  by observing the appearance of the electron neutrinos in the far detector. In practice,  to observe electron neutrinos the electrons produced in charged current $\nu_e$-target interactions must be observed. Simultaneously the detector will be able to detect $\pi^0$s which are produced in neutral current neutrino-target interactions. This is important  background for $\nu_e$ measurement because $\pi^0$ decays into two photons which can be misled  with the electron shower. Since a small number of electron neutrinos  is expected in the far detector,  success of the experiment may depend on proper estimation of the number of $\pi^0$s produced in neutral current neutrino interactions.

The description of $\pi^0$ production in  neutral current neutrino-nucleon scattering still requires  careful attention. As has been already mentioned  there are limited experimental data for these processes. Standard model gives a hint on how to construct the hadronic current for neutral current interaction. The neutral current (NC) is expressed thought the third isovector component of the vector-axial current and the electromagetic current
\begin{equation}
\mathcal{J}^{NC} = \mathcal{J}^{CC,I_3} -  2\sin^2\theta_W\,
\mathcal{J}^{EM},
\end{equation}
where $\theta_W$ is the Weinberg angle. The relative weight of the vector and axial contributions  to the neutral current cross sections is different than in the case of the charged current reactions. One can imagine a model which fits well to charged current data, but  overestimates axial and underestimates vector contributions. In that case predictions for neutral current reactions will be wrong.

In this talk we will present comparison of two descriptions for single pion production in neutrino-nucleon scattering, namely the Rein Sehgal (RS) model \cite{Rein:1980wg} and the isobar formalism \cite{Schreiner:1973mj}. The first model is implemented in the Monte Carlo codes which were/are used in data analysis of K2K and MiniBooNE experiments. We will  discuss how to improve the Rein Sehgal model to get more precise description of the charged current and neutral current cross sections (in  particular for $\pi^0$ production). Our  idea is to consider new vector and axial form factors. We  express the RS model form factors by the ones from the isobar formalism.  The new vector contribution fits well to the electoroproduction data, while the axial contribution was obtained in  simultaneous  fit to the ANL and BNL differential cross section $d\sigma/d Q^2$  data.

\section{Single pion production}

\begin{figure}\centering{
\includegraphics[width=\textwidth]{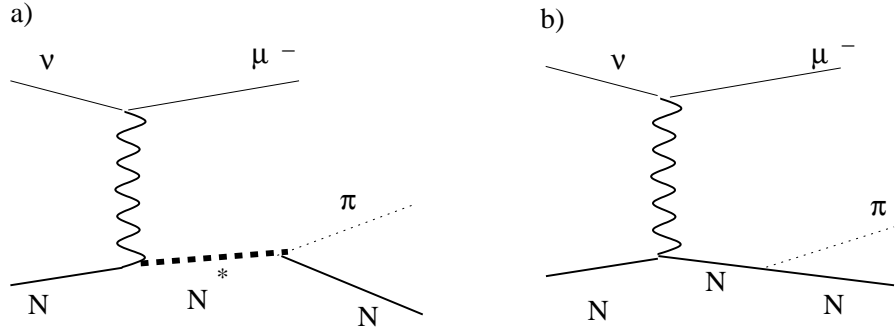}}
\caption{In Fig. (a) Feynman diagram for the resonance pion production in $\nu N$ scattering is shown. In Fig. (b) one of the typical Born diagrams for the nonresonant pion production in $\nu N$ scattering is presented. \label{Fig_process}}
\end{figure}

One can distinguish two mechanisms for single pion production in neutrino-nucleon scattering:
\begin{enumerate}
 \item[(i)] resonant  -- the nucleus is excited to the resonance state, then the resonance decays to a single pion and nucleon. It is illustrated  in Fig. \ref{Fig_process}a);
\item[(ii)] nonresonant -- the pion is produced without resonant excitation of the nucleon. It can be described by Born diagrams (Fig. \ref{Fig_process}b)) \cite{Fogli:1979cz}.
\end{enumerate}
Since the resonance is not directly observed (only its decay products), it is impossible to distinguish between both mechanisms.

However,  it seems that for the reaction:
\begin{equation}
\label{reaction}
\nu + p  \to \mu^- + \pi^+ + p.
\end{equation}
the nonresonant contribution is negligible -- to fit   the hadronic invariant mass distribution of the $\pi^+ p$ system it is enough to consider   the Breit-Wigner distribution of the $P_{33}(1232)$\footnote{It is the 3/2 spin resonance denoted also by $\Delta(1232)$.} resonance  \cite{Kitagaki:1986ct}. Therefore process (\ref{reaction}) seems to be  appropriate for fine tuning the description of the resonance contribution to the pion production in $\nu N$ scattering. For other possible channels the nonresonant contribution  as well as heavier than 1232 MeV resonances must be taken into account. However, if the neutrino has energy around 0.7 GeV (characteristic energy for T2K project) it will be enough to consider only  the $P_{33}(1232)$ resonance.

In this talk we present results of comparison between the RS model and the isobar formalism for the $P_{33}(1232)$  excitation. Both descriptions should lead to  similar resonance helicity amplitudes for the $P_{33}(1232)$ excitation.

\subsection{Rein Sehgal model}

The Rein Sehgal model is an extension of the relativistic harmonic oscillator quark model (FKR -- Feynman, Kislinger and Ravndal)  \cite{FKR}. The FKR model was applied to describe photo- and electro- production of resonances \cite{Ravndal71}, and then neutrino-production of resonances \cite{Ravndal72},\cite{Ravndal73}.

In the FKR model baryon wave functions are given by  symmetric representations of $SU_{flavor}(2) \times SU_{spin}(2)\times O(3)$. The nucleon and $\Delta(1232)$ are ground states of the harmonic oscillator. Other excited states of the nucleon are classified  by:  level in the harmonic oscillator model, spin, orbital angular momentum etc..  Feynman \textit{et al.} proposed  an operational form of one body quark currents for electromagnetic and weak transition of nucleon to resonance state. Knowing  the nucleon and resonance  wave functions  one can compute the following  helicity amplitudes:
\begin{eqnarray}
\label{helicity_amplitudes_1}
f_{\underline{0}}&=&\left<N^*, s'\right| \mathcal{J}_{t} + \frac{\nu_{res}}{q_{res}}\mathcal{J}_{z}\left| N,s \right>, \\
\label{helicity_amplitudes_2}
f_\pm &=&\left<N^*, s' \pm 1\right| \mathcal{J}_\pm\left| N,s \right>, \quad\mathcal{J}_\pm =\mp\frac{1}{\sqrt{2}}\left(\mathcal{J}_x \pm \mathrm{i}\mathcal{J}_y \right),
\end{eqnarray}
where $Q^2= -q_\mu^2$, $q^{\mu} = (\nu_{res},0,0,q_{res})$ and $\nu_{res}$, $q_{res}$ are energy, momentum transfers computed in the resonance rest frame, $s$ and $s'$  denote spins of incoming and outgoing particles.

The hadronic charged electro-weak current has vector -- axial structure: $\mathcal{J}^{CC}= \mathcal{J}^V - \mathcal{J}^{A}$. In the formalism proposed by Feynman \textit{et al.} the vector and axial one body quark currents are multiplied by two phenomenological functions: the vector and axial form factors
$$\mathcal{J}^{V,A} \to G_{V,A}\mathcal{J}^{V,A}.$$
Final hadronic state can be either a nucleon or a resonance. It means that formalism should be able to reconstruct helicity amplitudes for elastic $ep$ scattering and quasi-elastic $\nu n$ scattering. This property was applied to find  expressions for  $G_{V,A}$. They are proportional to the elastic vector and axial nucleon form factors:
\begin{eqnarray}
G_V(Q^2) & =&  \left(1 + \frac{Q^2}{M_V^2} \right)^{-2}\left(1 + \frac{Q^2}{4 M^2}\right)^{\frac{1}{2}} \\
 G_A(Q^2) & =& \left(1 + \frac{Q^2}{M_A^2} \right)^{-2}\left(1 + \frac{Q^2}{4 M^2}\right)^{\frac{1}{2}},
\end{eqnarray}
where $M_V$ and $M_A$ are vector and axial masses while $M$ denotes nucleon mass.

Rein and Sehgal extended the FKR model to describe single pion production induced by charged and neutral currents neutrino-nucleon interactions. Pions are produced by resonance excitations but the description of the nonresonant  contribution was also proposed. In the RS approach 18 resonances with hadronic invariant mass smaller than 2 GeV are discussed.
\begin{figure}
\centering{
\includegraphics[width=\textwidth, height=5.5cm]{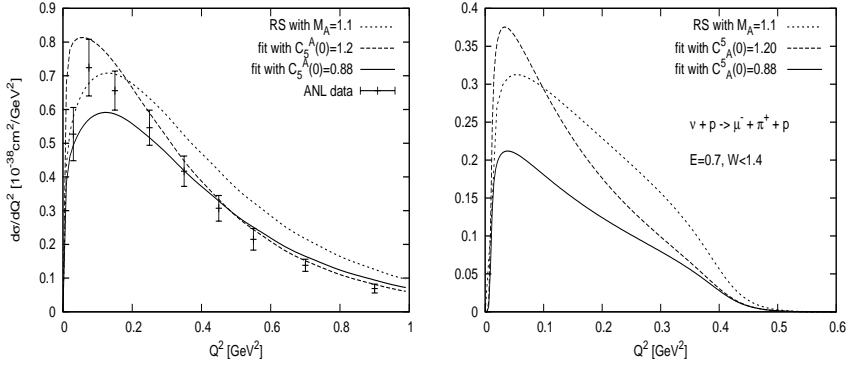}}
\caption{ $d\sigma/dQ^2$ differential cross sections for reaction $\nu + p \to \mu^-  + \pi^+ + p$.
In the left figure cross sections computed for the ANL beam are presented \cite{Radecky:1981fn}. In the right figure the differential cross sections are computed for $E=0.7$~GeV -- an average neutrino energy for T2K experiment.
The cross sections are computed with the original RS form factors (dotted lines), and our fits. The solid and dotted lines denote cross sections obtained with (\ref{c5A_fitbez}) and (\ref{c5A_fit12}) axial form factors respectively. The cut on the invariant hadronic mass $W<1.4$~GeV is
imposed.\label{Fig_ANL}}
\end{figure}

It must be emphasized that the RS model successfully described neutrino cross section data if some reasonable accuracy was assumed. However, higher precision measurements require  more detailed description of the cross sections. We noticed that the RS model underestimates inclusive $F_2$ $ep$ structure function \cite{Graczyk:2007bc} and consequently $ep$ cross sections. Therefore we proposed an effective description of the vector contribution, which more accurately fits to the $ep$ experimental data.

As was mentioned above next generation long baseline experiments will measure neutrino-nucleus scattering mainly in the $\Delta(1232)$ resonance region. Therefore our improvements concern the $P_{33}(1232)$ resonance.

%%%%%%%%%%%%%%%%%%%%%%%%%%%%%%%%%%%

\subsection{Isobar formalism}

In the FKR model the axial and vector form factors were obtained by referring formalism to elastic and quasi-elastic scattering.  But  one can propose a different idea for introducing form factors into the model. Since  the $\Delta(1232)$ resonance region is of our interest,  it is natural to use knowledge about the $P_{33}(1232)$ production as reference point for the FKR/RS form factors. For that reason it is useful to compare the isobar formalism for the $P_{33}(1232)$ excitation \cite{Schreiner:1973mj} with the RS approach.

The isobar formalism gives  very phenomenological, only in terms of form factors, description for the $\Delta(1232)$ excitation. Helicity amplitudes computed with this approach can be easily compared with corresponding once in the RS model. Since the isobar model has been updated \cite{Lalakulich:2006sw} lastly,  the comparison may show a hint on how to improve the RS model predictions.

\begin{figure}
\centering{
\includegraphics[width=\textwidth, height=5.5cm]{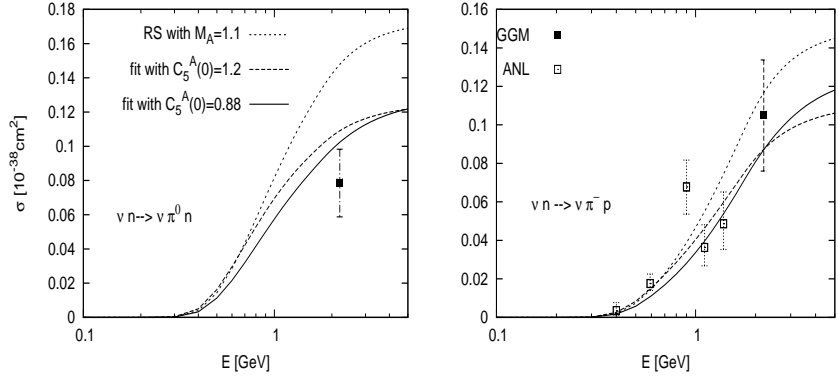}}
\caption{Total cross sections for two neutral current reactions: $\nu + n \to \nu + \pi^0 +n$ (left figure) and
$\nu + n \to \nu + \pi^- +p$ (right figure). The solid and dashed lines correspond to cross sections computed with
 (\ref{c5A_fitbez}) and (\ref{c5A_fit12}) axial form factors respectively, while by dotted lines cross sections obtained with the RS form factors ($M_A=1.1$~GeV) are shown.   The data is taken from Ref. \protect\cite{Krenz:1977sw} (black squares) and Ref.
\protect\cite{Derrick:1980nr} (white squares). The cut on the invariant hadronic mass $W<2$~GeV is
imposed.\label{Fig_cross_NC}}
\end{figure}

In the isobar formalism the hadronic current for the $P_{33}(1232)$ excitation is constructed to satisfy Lorentz invariance \cite{Schreiner:1973mj}.
The initial  nucleon state is given by Dirac spinor while  the $P_{33}(1232)$ state is the Rarita-Schwinger field. The vector current is conserved and satisfies CVC hypothesis while  axial current is constrained by PCAC hypothesis. These assumptions reduce the number of independent form factors.

The vector current for $\Delta^{++}(1232)$ excitation is expressed by three form factors $C_3^V$, $C_4^V$ and $C_5^V$
\begin{equation}
\left<\Delta^{++}(p') \right| \mathcal{J}_\mu^{V} \left| N(p)
\right> = \sqrt{3} \bar{\Psi}_\lambda (p') \left[ g^{\lambda}_{\
\mu} T_\nu q^\nu - q^\lambda T_\mu \right]\gamma_5 u(p),
\end{equation}
where
\begin{equation}
T_\mu =\frac{C_3^V}{M}\gamma_\mu + \frac{C_4^V}{M^2}p'_\mu +
\frac{C_5^V}{M^2}p_\mu,
\end{equation}
and $g_{\mu\nu}$ is Minkowski metric.

Axial current depends also on  three form factors $C_4^A$, $C_5^A$ and $C_6^A$
\begin{equation}
\left<\Delta^{++} (p')\right|\mathcal{J}_{\mu}^{A}
\left|N(p)\right> = \sqrt{3}\bar{\Psi}_\lambda (p') \left[
g^{\lambda}_{\ \mu} B_\nu q^\nu - q^\lambda B_\mu + g^{\lambda}_{\
\mu}C_5^A + \frac{q^\lambda q_\mu}{M^2} C_6^A\right]u(p),
\end{equation}
where
\begin{equation} B_\lambda = \frac{C_4^A}{M^2}p'_\lambda.
\end{equation}
From the PCAC hypothesis
$$ C_6^A(Q^2) =   \frac{M^2}{m_\pi^2 + Q^2} C_5^A(Q^2),$$
where $m_\pi$ is pion mass. Usually the Adler relation
$ C_4^A = - C_5^A/4$
is also assumed. Then the axial current depends only on  $C_5^A(Q^2)$ form factor.

\section{Comparison of RS model with isobar formalism}

To compare the RS model with isobar formalism three independent  helicity amplitudes (see expressions: \ref{helicity_amplitudes_1}-\ref{helicity_amplitudes_2}) are evaluated in both approaches. It is well known that the quark model is not able to reproduce a charged contribution of the $P_{33}(1232)$ resonance (expression (\ref{helicity_amplitudes_1}) vanishes). Therefore, the exact comparison is impossible but  it is still reasonable because the contribution from  helicity amplitude (\ref{helicity_amplitudes_1}) is very small. Comparing helicity amplitudes allows us to express $G_V$ by $C_3^V$, $C_4^V$ and $C_5^V$ functions (for more details see Ref. \cite{Graczyk:2007bc}):
\begin{equation}
\label{G_V_general_fit} G_V^{new}(W,Q^2) = \frac{1}{2}
\sqrt{  3\left(G^{f_{3}}_{V}(W,Q^2)\right)^2 +
\left(G^{f_1}_{V}(W,Q^2)\right)^2 },
\end{equation}
where
\begin{eqnarray}
\label{G3V} G_{V}^{f_3}(W,Q^2) & \equiv & \frac{F(Q^2, W)}{2\sqrt{3}}
\left[C_4^V
\frac{W^2-Q^2-M^2}{2 M^2} \right. \nonumber \\
& & \;\;\;\;\;\;\;\;\;\left.+ C_5^V\frac{W^2+Q^2-M^2}{2M^2} +\frac{C_3^V}{M}(W + M) \right],\\
\label{G1V} G_{V}^{f_1}(W,Q^2) & \equiv
&-\frac{F(Q^2, W)}{2\sqrt{3}}
\left[C_4^V\frac{W^2-Q^2-M^2}{2 M^2}\right. \nonumber\\
& & \left. \;\;\;\;\;\; +
C_5^V\frac{W^2+Q^2-M^2}{2M^2} - C_3^V\frac{(M + W)M +Q^2}{M W}
\right]. \\
F(Q^2, W) & = & \left(1 +
\frac{Q^2}{(M+W)^2} \right)^{\frac{1}{2}},
\end{eqnarray}
where by $W$ hadronic invariant mass is denoted.

Lalakulich \textit{et al.} \cite{Lalakulich:2006sw} did some effort to fit $C_3^V$, $C_4^V$ and $C_5^V$ form factors to resonance photo-production data. The obtained parameterizations lead to a vector contribution which agrees with elastic $ep$ experimental data, and with MAID predictions \cite{Drechsel:1998hk}. It seems that a use of these form  factors  guarantees adequate description of the resonance vector contribution in neutrino-nucleon scattering. Therefore we express  $G_V$ form factor by Lalakulich \textit{et al.} parameterizations.

Similarly as in the case of vector contribution three independent helicity amplitudes are computed for axial current. Here, the relations between amplitudes are more complicated than in the vector current case and it was possible only to find an approximate relation between $G_A$ and $C_5^A$:
\begin{equation}
\widetilde G_A^{new}(W,Q^2) = \frac{\sqrt{3}}{2} F(Q^2, W) \left[1 -\frac{W^2 -Q^2
-M^2}{8M^2} \right] C_5^A(Q^2),
\end{equation}
where, in our notation $\widetilde{G}_A = (0.76)\cdot G_A$.

To get a proper form of the $C_5^A(Q^2)$ axial form factor we proposed a numerical procedure \cite{Graczyk:2007bc} which allows us to find  simultaneous fit to both ANL and BNL data sets. We fixed ANL cross sections normalization then the normalization of the BNL data had to be changed in order to obtain a consistent fit.

We discussed two different parameterizations of the axial form factor.

In the first case we assumed that $C_5^A(0)$ is constrained by PCAC, and equals 1.2. For that case we obtained:
\begin{equation}
\label{c5A_fit12} C_5^A(Q^2) = \frac{C_5^A(0)}{\left( 1 +
\displaystyle\frac{Q^2}{M_a^2}\right)^2 },
\end{equation}
with axial mass $M_a^2 = 0.54$~GeV$^2$.

We consider also parametrization with $C_5^A(0)$ treated as a fit parameter. In that case we assumed that $C_5^A$ has functional form:
\begin{equation}
\label{c5A_fitbez} C_5^A(Q^2) = \displaystyle \frac{C_5^A(0)}{\left(
1 + \displaystyle \frac{Q^2}{M_a^2}\right)^2 \left( 1 +
\displaystyle \frac{Q^2}{M_b^2}\right)}.
\end{equation}
We got $C_5^A(0) \approx 0.88$,
with $M_a^2 \approx 9.71$  GeV$^2$ and  $M_b^2 \approx
0.35$ GeV$^2$. It is interesting to notice that the value of  $C_5^A$ at $Q^2=0$ is very similar to $C_5^A(0)=0.867$ which was computed in more advanced approach (chiral constituent quark model) \cite{BarquillaCano:2007yk}.

\section{Numerical results}

In this section we show how the new form factors, described in the previous section,  change differential and total cross sections for neutrino-nucleon scattering.

In Fig. \ref{Fig_ANL} we plot cross sections computed with new form factors and the RS model form factors. In the left panel of Fig. \ref{Fig_ANL} theoretical predictions are compared with $d\sigma/ dQ^2$ ANL data for reaction $\nu + p \to \mu^- + \pi^+ + p$.   The cross section computed with the axial form factor (\ref{c5A_fit12}) fits better to ANL data than the fit given by Eq. (\ref{c5A_fitbez}).  Small value of  axial form factor (\ref{c5A_fitbez}) at $Q^2=0$ ($C_5^A(0)=0.88$) leads to significant reduction of $d\sigma/d Q^2$ cross section at low $Q^2$. Let us also remark that this form factor fits better to the BNL $d\sigma/ d Q^2$ data  (see Fig. 3 in Ref. \cite{Graczyk:2007bc}).

In the right panel of Fig. \ref{Fig_ANL} we show the differential cross sections computed for neutrino energies E=0.7 GeV -- an expected averaged energy of  T2K neutrino beam. Similarly as in the left panel, we present cross sections  for charged current reaction $\nu+p\to\mu^- + \pi^+  + p$. We see that three different parameterizations of the axial form factor lead to  different $Q^2$ dependence of differential cross section. Let us notice that the the original the RS  axial form factor is a kind of compromise between the axial form factors preferred by ANL and BNL data. From that point of view it seems that a use of the Rein Sehgal description leads to  world averaged values of neutrino cross sections.

It has been already mentioned that the cross sections for pion production in neutral current neutrino-nucleon scattering are of extreme interest.  In Fig. \ref{Fig_cross_NC} we plot total cross sections for two neutral current reactions:  $\nu + n \to \nu + \pi^0  +n$  and $\nu + n \to \nu + \pi^- + p$. The cross sections computed with the form factors presented in this talk  are by 20\% smaller than predictions obtained with  the original RS model (with axial mass 1.1 GeV). However, the small number of  data points did not allow  to verify which approach fits better to the data.

\section{Final remarks}

In this talk we presented some improvements of the RS model. We proposed modifications of the RS description which correct description of the vector and axial contributions. Our improvements can be  easily applied to  existing Monte Carlo codes. We believe that after our modifications the RS model more precisely describes total and differential cross sections.

We showed also that it is possible to find a simultaneous fit to both ANL and BNL data, but the normalization of  BNL data had to be changed by about 25\%.

One of our parameterizations of the axial form factor (\ref{c5A_fitbez}) leads to the $d\sigma/dQ^2$ differential cross section which is significantly reduced at small $Q^2$. It  is the parametrization preferred by BNL data. It is possible that a use of this form factor in the data analysis may help to understand the observed at K2K and MiniBooNE discrepancy between experimental data and theoretical predictions at low $Q^2$.

\end{document}